# Crowd simulation influenced by agent's socio-psychological state

F. Cherif, and R. Chighoub

**Abstract** — The aim our work is to create virtual humans as intelligent entities, which includes approximate the maximum as possible the virtual agent animation to the natural human behavior. In order to accomplish this task, our agent must be capable to interact with the environment, interacting with objects and other agents. The virtual agent needs to act as real person, so he should be capable to extract semantic information from the geometric model of the world where he is inserted, based on his own perception, and he realizes his own decision. The movement of the individuals is representing by the combination of two approaches of movement which are, the social force model and the based-rule model. These movements are influenced by a set of socio-psychological rules to give a more realistic result.

——————————— ◆ ———————————

## 1 INTRODUCTION

VIRTUAL human simulations are becoming each time more popular, and many systems are available targeting several domains, as autonomous agents, humans factors analysis, training, education, virtual prototype simulation-based design, and entertainment simulations with autonomous virtual humans, or actors, may use different techniques for the behavioral programming. Most common approaches are based on scripts and hierarchical finite state machines, but many other techniques exist, as the parallel transitions network. Some constraints arise when we deal with crowds of virtual actors different from the modeling of virtual individuals; crowds are ubiquitous feature of everyday life. Human crowds are ubiquitous in the real word, making their simulation a necessity for realistic interactive environments, physically correct crowd models also have applications outside of computer graphics in psychology, transportation research, and architecture.

People have long assembled collectively to observe, to celebrate, or to protest at various happenings. The collective assemblages or gatherings called crowds have been the object of scientific inquiry since the end of 19th century. With computers it becomes possible not only to observe human crowds in the real world, but also to simulate various phenomena from the domain of collective behavior in the virtual environments. Collective behaviors have been studied and modeled with very different purposes. Besides single work concerned with generic crowd simulation, most approaches were application specific, focusing on different aspects of the crowd behavior. As a consequence they employ different modeling techniques ranging from those that do not distinguish individuals such as flow and network models, to those that represent each individual as being controlled by rules based on physical laws or behavioral models. Applications include animation production systems used in entertainment industry, crowd behavior models used in training of military personnel or policemen, crowd motion simulations to support architectural design both for everyday use and for emergency evacuation conditions, simulations of physical aspects of crowd dynamics and finally sociological and behavioral simulations.

Within the framework of our work, we propose a microscopic model of simulation of a virtual crowd with high density in a dynamic and more complex environment. In order to navigate in a complex environment, we need to have an efficient abstract representation of the virtual environment [4] where the agents can rapidly perform way-finding. For this reason, we use two different approaches for represent our space, which are cell graphs and portal graph. These two approaches of abstract representation can also be used to store some pre-computed information about the environment that will speed up the navigation and also be helpful to achieve fast perception for local motion computation. This virtual environment is similar to our real world, filled with an important number of intelligent entities (e.g. virtual agents, autonomous agents, intelligent objects). The aim of our work is to create virtual humans as intelligent entities in these space, which includes approximate the maximum as possible the virtual agent animation to the natural human behavior. In order to accomplish this task, our agent must be capable to interact with the environment, interacting with objects and other agents. The virtual agent needs to act as real person, so he should be capable to extract semantic information from the geometric model of the world where he is inserted, based on his own perception, and he realizes his own decision. The movement of the individuals is representing by the combination of two approaches of movement which are, the force social model and the rule-based model. These movements are influenced by a set of socio-psychological rules to give a more realistic result.

## 2 RELATED WORK

————————

- *Chighoub Rabiaa, Computer science department, LESIA, Biska University, BP 145, Biskra 07000.*
- *Cherif Foudil, Computer science department, LESIA, Biska University, BP 145, Biskra 07000.*

Computers power increase recently allowed to populate interactive virtual worlds [9] with numerous inhabitants [8]. Crowds are now common in movies and more and more in video games, especially in real-time strategy games. Simulation of real-time virtual crowds is still a difficult challenge given that available computation-time is mainly dedicated to rendering; a need for fast simulation techniques exists. The topic of modeling of pedestrian streams is not new and has been done for many years. The previous focus was on modeling pedestrian streams in urban environments. In these earlier studies, the aim was to determine the dimension of the parameters of walkways. Later the scope was extended to the field of emergency. In the 90's the simulation of pedestrian streams was integrated in the simulation of intermodal transport facilities. Nowadays, there exist quite a three different approaches for modeling pedestrian streams.

1. Macroscopic simulation: Models of this category are field-based simulation models that only deal with densities and flux. The approach describes pedestrian flows with differential equations, and is based on the idea that the movement of pedestrians can be handled analogous to fluids and gases.
2. Microscopic simulations: these models consider the movement of pedestrians at the individual level [1]. They treat pedestrians as individuals and model the individual's behavior with the expectation that dynamic crowd behaviors will emerge from the interactions between individuals and the environment.
3. Mesoscopic simulation: Models of this approach are a mixture of macroscopic and microscopic models, simulating not individual entities but groups of similar entities, The idea of grouping individuals was transformed to mesoscopic pedestrian flow model. That means we do not model a single pedestrian but we use groups of pedestrians and every group has its own rules of behavior. The flow of a single pedestrian is integrated into a flow of pedestrian groups. This simplification is valid because one of the main interesting results from the simulation model is not the state of a single person but the number of persons in a particular area at time t. Figure 1 shows the classification of the mesoscopic approach in relation to the other approaches.

In our work, we interest with microscopic approach, many microscopic simulations already exist, and one way to group them is to divide them into cellular automata models, behavioral force models, and rule-based models. The key difference between these techniques is the method in which the entity is represented, how the entity is controlled, and how it interacts with other entities.

The social forces model [5, 6] calculates forces acting on agents to determine movement, with excessive forces leading to agent injuries. The model considers the effect that each agent has upon all the other agents, almost as if the model were a simulation of an n-body problem in astrophysics. Physical forces (e.g. friction encountered when brushing past another person, or elastic force due to body compressions) are modeled, as are social forces (desire to change direction to avoid another). One problem with Helbing's model is that of computational complexity. Simulation update is $O(n^2)$ due to the calculation of the effect that each agent (and obstacle) has on all the other agents. This may limit the model's ability to simulate many agents. Braun et al [2] expand Helbing's social force model. In their model, each pedestrian is assigned a 'family' identifier and an 'altruism' level. These acts as forces in the model to tend to make some of the pedestrians form groups with others.

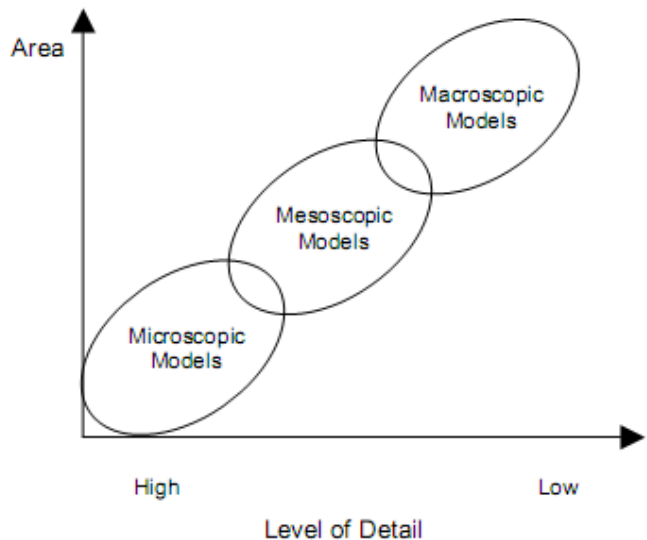

Fig. 1. Relations of three approaches of modeling pedestrian

Reynolds [15] described a distributed behavior model for simulating flocks of birds formed by actors endowed with perception skills. In fact, the birds (or 'boids') maintain proper position and orientation within the flock by balancing their desire to avoid collisions with neighbors, to match the velocity of neighbors and to move towards the center of the flock. Reynolds's work shows realistic animation of groups by applying simple local rules within the flock structure. Following this approach, Terzopoulos et al. [20] presented the simulation of virtual fish groups, where each fish is endowed with perception (artificial vision), locomotion control (based on a mass-spring system used to propel the fish in the water) and with behaviors (based on a group of parameters, such as hunger degree and predators fear).

## 3 ARCHITECTURE OVERVIEW

We have developed a realistic simulator of high-density crowds of autonomous virtual agents composing of an important number of pedestrians autonomous in reconstructed large environments, demonstrating realistic human activity. In this context, a microscopic architecture has been proposed for handling high-density crowds of autonomous agents moving in a natural manner in dynamically changing virtual environments filled with obstacles.

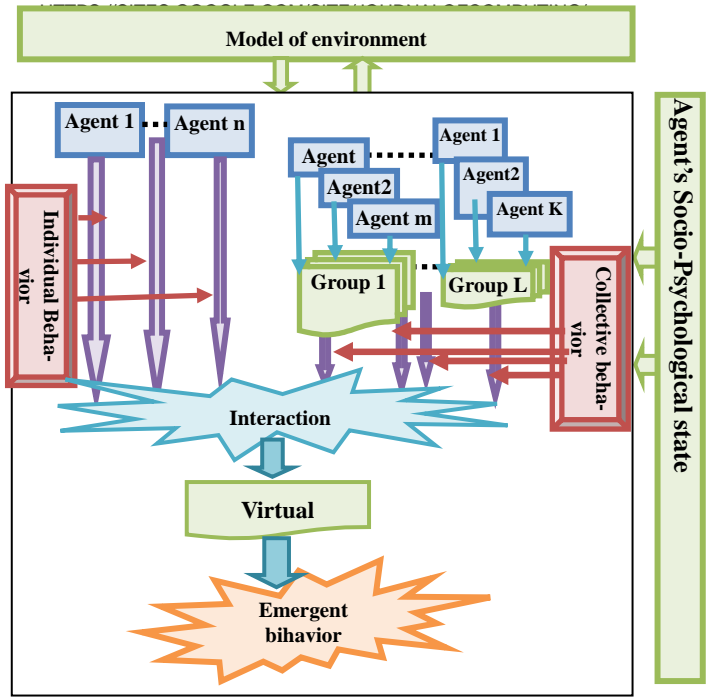

Fig. 2. Our system architecture

In this architecture (proposal), a virtual agent can be defined as an autonomous entity in a virtual environment. It should not only look like, but also behave as a living organism in a synthetic 3D world, and be able to interact with the world and its inhabitants. The virtual agent must receive its environment through sensors or perception, it receives from the environment the list of entities (agents, obstacles…) that are in its field of view, so he should be capable to extract semantic information from the geometric model of the world where he is inserted, and based on his own perception, he decides its decision and acts as real people. The complexity of pedestrian behavior comes from the presence of collective behavioral patterns (as clustering, lanes and queues) evolving from the interactions among a large number of individuals. This empirical evidence leads to consider two different approaches: pedestrians as a flow and pedestrians as a set of individuals or agents.

### 3.1 Model of environment

In any location, and particularly in a city, virtual humans need to be aware of their environment in order for them to avoid any collision, know where their goals are, and how to reach them. Many different approaches have been developed to tag an environment with information. Our system can handle two approaches of environment representation, grid and topological graph, by combining both paradigms – grid-based and topological—the approach presented here gains the best of both worlds: accuracy/consistency and efficiency.

1. Grid-based approach, represent environments by evenly spaced grids [18, 17]. Each grid cell may, for example, indicate the presence of an obstacle in the corresponding region of the environment. This approach must be fine enough to capture every important detail of the world [7, 19].
2. Topological approaches represent environments by graphs. Nodes in such graphs correspond to distinct situations, places, or landmarks (such as doorways) [19]. They are connected by arcs if there are direct paths between them. [18, 17]. Consequently, they permit fast planning.

The process of representation of the environment needs:
1. Our system receives as an input an arbitrary building model in 3 dimensions and creates grid map. To build the grid graph, by cutting out the environment a unit of cells, then; one encapsulates all information (static) necessary to the process of navigation.
2. Once we have the grid decomposition, we start an iterative conquering process starting from the top left corner cell that is empty. We assign a positive number to this cell that will represent the room ID in the cell and portal graph, and then this ID is propagated using a breadth-first traversal. The propagation of the cell ID continues until the entire room is bounded by cells having either 0 (wall) or -1 (door) [10].

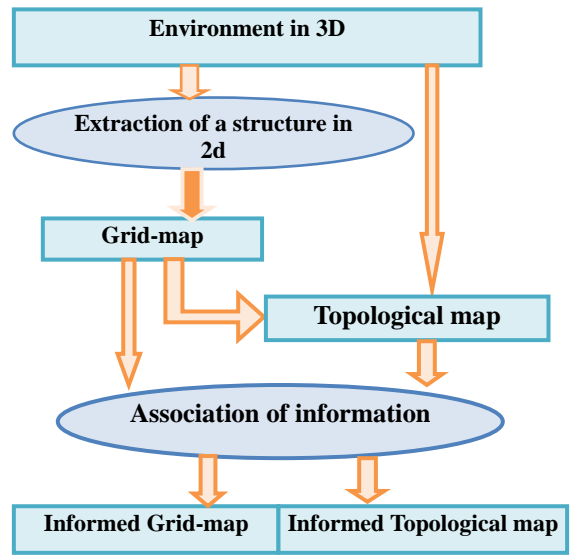

Fig. 3. Model of environment

3. Once all the cells have been identified, we need to generate the cell and portal graph by joining the rooms through the doors. This is carried out by traversing the grid representation from left to right, top to bottom, looking for doors. When a door is found, a portal is created that will join the two cells appearing at both sides of the door [11].
4. In the last step, points of attraction will be associated at each door and at each painting. In order to perform global navigation, we also need to store information about paths within the building from each cell to each of the exits in the building.

Each cell will contain one or more alternative paths to each exit [11].

## 3.2 Model of agent

Our simulation consists of high-density crowd of autonomous virtual human agents existing in dynamic, complex, virtual 3d environment. In order to behave in believable way these agents must act in accordance with their surrounding environment, be able to react to its changes, to the other agents and also to the actions of real humans interacting with the virtual world. Figure 4 gives overview of the agent model, this model of the agent follow the Sense-Decide-Act (SDA) cycle, where it can be broken up into subsystems, each one associated in turn with more specific routines. From the perception submodule, the individual perceive the environment according to its position, personality and knowledge, and he obtains the environment information on a semicircular front region of the agent, it detects the positions, orientations and speeds of other agents, and obstacles. Moreover, the agent cans predicted future positions changes in speed and orientation of each character and of objects in the VE.

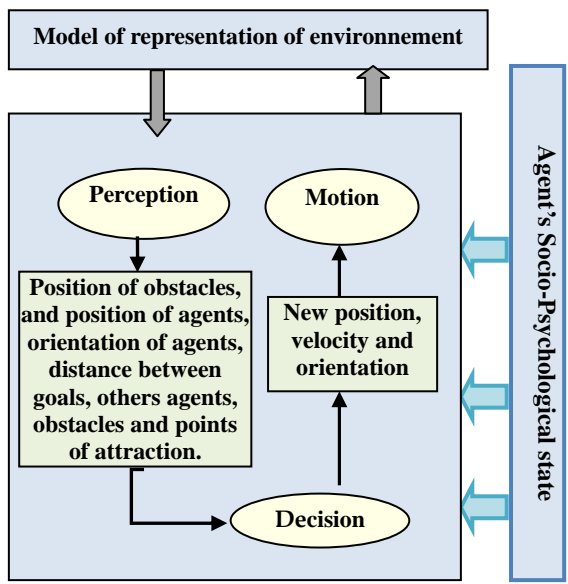

Fig.4. Model of agent

Based on information perceived, the internal state and the socio-psychological traits of the agent (current behavior, panic, role, impatience…), the Decision sub-model calculates the orientation, the velocity and next position of the agent, finally the Motion sub-model execute and realize this decision.

### 3.2.1 Perception

Humans can perceive a field of view (FOV) from 120° to 180°, the latter being the most common we can simulate human perception by having the virtual agents be only aware of those objects falling within a specified angle from their direction of movement (assuming the head is oriented in the same direction). This is calculated from the dot product between the direction of movement vector and the vector joining the current position with each object in the room. Since the dot product gives us the cosi that value is bigger than 0 it means the object falls within the agent's FOV [21]. In figure 5, the agent perceives its environment; and determines which object exists in its zone of vision, wall, the obstacle and another agent J.

### 3.2.2 Motion

The main motivation for this model is the fact that different people can react in different ways depending on their individual characteristics and on their psychological traits. Pedestrian movement shows the characteristics of a pedestrian compared to cars or other vehicles such as a bicycle, a pedestrian has more flexibility to move in two dimensions continue space, a pedestrian also has more flexibility to stop and go and a pedestrian also has the smallest velocity and the highest fluctuation in acceleration and velocity compared to vehicles. In our system, which consists in simulating a dense human crowd, the local movements of agents are inspired by the combination of two approaches which are the approach based-rules and the approach of social force. Reynolds [15, 16] described a distributed behavioral model for simulating flocks of birds formed by actors endowed with perception skills with many purposes. In fact, the birds (or boids) maintain proper position and orientation within the flock by balancing their desire to avoid collisions with neighbors, to match the velocity of neighbors and to move toward the center of the flock. Reynolds work shows realistic animation of groups by applying simple local rules within the flock structure. This model gives results realize for a crowd with low density. To mitigate the insufficiencies of the approach containing rules, we combined this approach with the approach of social force inspired of the model of Helbing [5]. It is based on physics and the socio-psychological forces in order to describe the human behavior of crowd in situations of panic, this model can simulate a crowd with high density but does not give realistic results. The displacement of the agents is modeled by a whole of social forces, and then Nuria [12, 13] hears this model by the incorporation of psychological and geometrical rules.

Therefore the module of movement of our pedestrian consists in calculating the new position and the new for each agent by using a certain force of attraction and repulsion, these forces make it possible the agent to show a broad variety of behaviors in an individual or collective way.

1. Advance Force: in the normal situation and in the absence of other pedestrians, the movement of a pedestrian should be directed from the current location P(t) toward the destination point E(t), there should be a "advance force" that directs the pedestrian to move, the advance force makes the pedestrian path almost in a straight line. So the agent perceives the environment and if its zone of vision does not contain obstacles or pedestrians, then it advances easily towards its goal by using its stable speed.

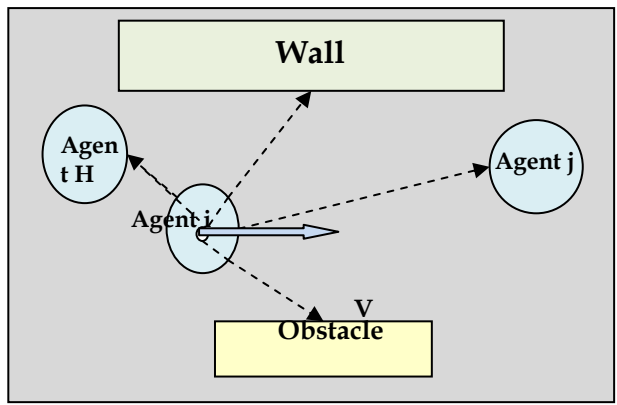

Fig.5. Process of perception

2. Acceleration Force: this behavior is influenced by the existence of panic, in this case if the zone of vision of our agent is empty, contains neither obstacles nor other virtual humans, the agent augments its speed, and is directed towards the door by using a maximum speed.

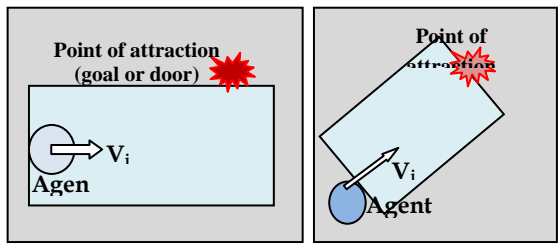

Fig.6. The Goal-attractive Force

3. The Occupant-attractive Force: because of the complex relationship among people, the cluster phenomena are observed, otherwise, people in danger tend to move towards the area with a large population, when the "follow" phenomena are often observed, which means occupants tend to follow the movement of majority. In the model, the occupant-attractive force is considered between some specified occupants when they are sufficiently far from each other (the distance between any two of them is no less than r1, and occupants may choose the same moving direction as most people have been taken around him (within a distance r2).

4. The Goal-attractive Force: In normal situation, pedestrians are sometimes attracted by window displays, sights, special performances (street artists), or unusual events at specified places (in certain area), otherwise, in the process of evacuation, occupants are attracted by the doors (in our model, the doors are marked by points of attractions). Both situations can be modeled by (often temporally decaying) attractive forces, in a similar way like effects repulsive, but with an opposite sign and a longer range of the interactions, it is generally assigned according to exits' quite selected in the building. The magnitude of attraction force is influenced by the type of situation (panic or normal), and the distance between the agent and the door (point of attraction), if this distance is short, the force is large.

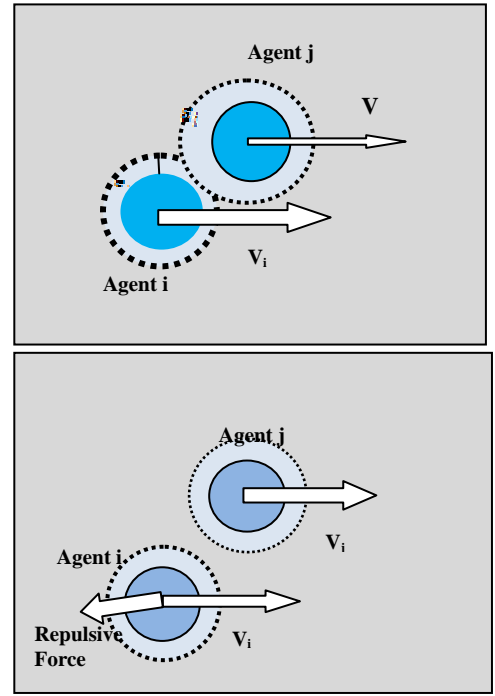

Fig.7. The Occupant-Repulsive Force

5. The Occupant-Repulsive Force: This repelling power works to avoid the collision between the pedestrians. To avoid the collision, it is supposed that each pedestrian has a ray of influence which represents its conscience of safety. The force is produced when the rays of influences of pedestrians overlap

6. The Obstacle-Repulsive Force: The response of collision, which is ensured by the repulsive forces, against the obstacles and the walls applies when the personal space of an agent covers with their zones. The covering of wall occurs when the distance between the center of the agent and the center of the wall is smaller than the personal space of the agent.

## 4 RESULTS AND PERFORMANCES

We have chosen the museum like an example of application of our system, the museum is regarded as a closed environment that its structure topology is represented by a unit of room of different size, connected between them by doors and passageways. Each room is generally to contain obstacles distributed on its surface, and it is decorated by a unit with the images suspended on the walls and by trimmings positioned in various localizations, in this environment our virtual agents, the normal situation, can move and see a number preset of the images and the trimmings which decorate the museum, but if the situa-tion changes, i.e. an alarm announces the event of panic, all the human virtual ones try to find the best way to leave the environment. Generally the pedestrian movements are influenced by a set of psychological and sociological rules which play a role of balancing of rationality in the behavior and the decision-making that we use to illustrate a broad more realistic variety of behaviors. Several situations have been studied to present the influence of socio-psychological rules in the realism of the behaviors human; we can resume the following situations:

1. The behavior of avoidance of collision and the avoidance of obstacles, these two behaviors are affected by the type of situation (normal and of panic), the personal space of the agent (broad, average, and narrow), and finally the level of patience of agent (patient and impatient).
2. Attractions forces, the agent are guided by three gravitational attractions which are Attraction force towards a point of attraction representing an under-goal (painting), Attraction force towards an exit (door), Attraction force towards another individual. This force is influenced by the distance between the individual and the goal (painting, doors or another agent), the distance which must respect between the agent and its goal, the situation (normal or of panic), knowledge of pedestrian (complete or partial), and the role of pedestrian (guide, or follower).

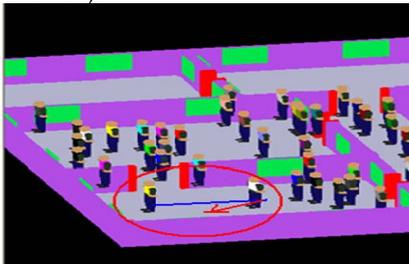

**The agent changes its direction
to avoid the collision**

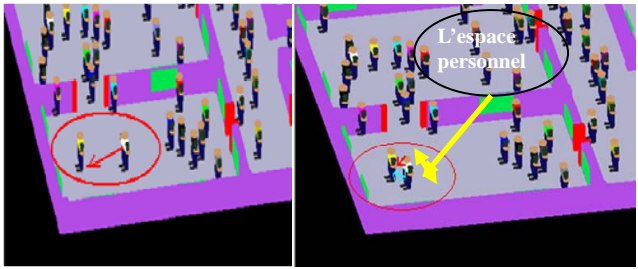

**The agent advances in its new direction, the change is rather fine since personal space is narrow**

(a) The avoidance of normal collision in case, such as personal space is narrow, and the impatient agent.

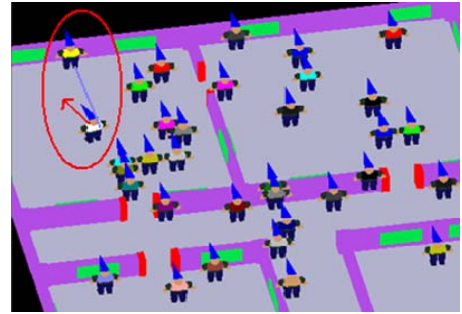

**Change of direction to avoid the collision**

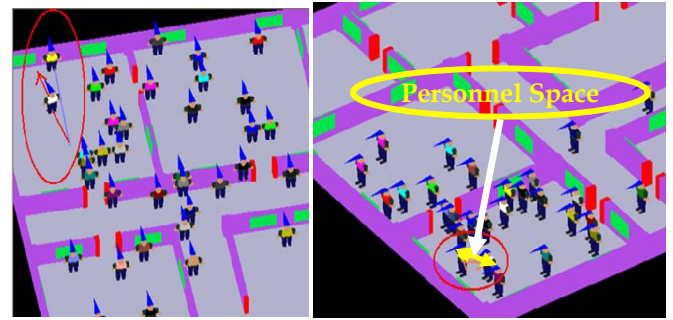

**Advance in the new direction**

(b) The avoidance of collision, such as personal space is average, and the impatient agent, the collision is treated with a broad distance (2 meter).

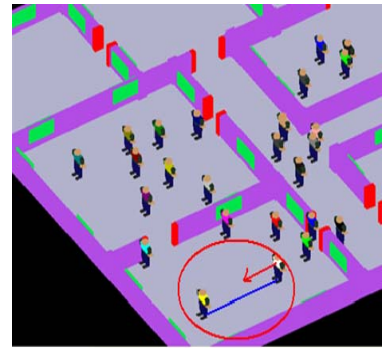

**A patient agent detects the collision, but it advances does not avoid it**

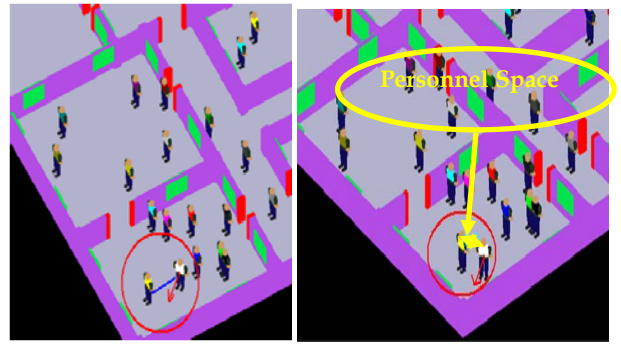

**The patient agent changes its direction at a short distance, and it advances in its new direction**

(c) The avoidance of collision, such as personal space is narrow, and the patient agent, the collision is treated with a short distance (1 meter).
Fig. 8. The behavior of avoidance of collision influenced by socio-psychological state

# 5 CONCLUSION

We have proposed a microscopic model of simulation of a crowd with high density of human vrtuels in a dynamic environment. We started with the description of an environment of simulation, not only from point of view topographic, but also from the semantic aspect. For this, we used the two approaches of representation of the environment knowing, the graph of grid and the topological graph.

The originality of our work is the introduction of a number of socio-psychological rules influencing the behavior of crowd. The integration of these rules in our model and the combination of the approach based rules and the approach of social force have conducted to realistic simulations.

Our system gives acceptable results, but optimizations and a whole of prospect remain to be supplemented like:
1. The addition of a cognitive layer in the human behaviors, the training in its process of navigation and the reasoning.
2. To integrate a responsible module to produce movements realistic, like walk, to run it, etc....
3. One can also introduce the concept of group and the behaviors collective, the agents that are very close can build groups and these groups are studied like only one entity, to minimize the computing time.

Dr. Cherif Foudil. is currently working as an Associate Professor of computer science at Computer Science Department, Biskra University, Algeria. Dr. Cherif holds PhD degree in computer science. The topic of his doctoral dissertation was Behavioral Animation: simulation of a crowd of virtual humans. He also possesses B. Sc. (engineer) in computer science from Constantine University 1985, M. Sc. in computer science from Bristol University, UK 1989. He is currently the head of "animation and artificial life" team in LESIA laboratory. His current research interest is in artificial intelligence, artificial life, crowd simulation, behavioural animation

**Chighoub Rabiaa.** A Phd candidate in computer science, her current research interest is crowd simulation.